# Harmonic hierarchy of altermagnetic spin splitting from symmetry-adapted wavefunctions

Yixuan Che,[1,*] Peibo Xu,[1,3,*] Haifeng Lv,[2,†] Xiaojun Wu,[2,3,‡] and Jinlong Yang[1,2,3]

[1]*Hefei National Research Center for Physical Sciences at the Microscale and School of Emerging Technology, University of Science and Technology of China, Hefei, Anhui 230026, China*

[2]*State Key Laboratory of Precision and Intelligent Chemistry, School of Chemistry and Materials Science, and Collaborative Innovation Center of Chemistry for Energy Materials (iChEM), University of Science and Technology of China, Hefei, Anhui 230026, China*

[3]*Hefei National Laboratory, University of Science and Technology of China, Hefei, Anhui 230088, China*

Altermagnets combine magnetic compensation with spin-momentum-locked splitting in the absence of spin-orbit coupling, yet existing descriptions, formulated primarily in terms of spin symmetry and lattice geometry, provide limited insight into the electric-structure perspective of its angular harmonic form. Here, we identify a wavefunction-level framework for altermagnetism in two-dimensional square lattices. Using symmetry-adapted polynomial wavefunctions, we show that the harmonic structure of momentum-space spin splitting is inherited from the geometry of the electronic wavefunctions which can be selected by crystal fields. Identical orbital sectors preserve conventional antiferromagnetic degeneracy, whereas intertwined linear and quadratic wavefunctions generate *d*-wave and *g*-wave altermagnetic anisotropies, respectively. Tight-binding analysis connects this hierarchy to inequivalent same-spin hopping channels. First-principles calculations on the *g*-wave **mcm**-type reticular material platforms confirm high-symmetry-linear degeneracy together with finite generic-*k* splitting. Our results establish a hierarchy linking wavefunction geometry, orbital realization, microscopic hopping anisotropy, and altermagnetic electronic structure.

*Introduction*. Altermagnetism has recently emerged as a distinct form of collinear magnetism that combines zero net magnetization with momentum-dependent spin splitting in the nonrelativistic limit [1-4]. This spin splitting distinguishes altermagnets from conventional antiferromagnets, where symmetry usually enforces spin degeneracy throughout the Brillouin zone. By allowing spin-polarized electronic states without macroscopic magnetization, altermagnets provide a platform for anomalous transport [5-7], topological phase transition [8,9], and the spin-current generation [10-12].

Current design principles for altermagnets are mainly formulated in terms of spin-group symmetry and crystallographic geometry [13-19]. In two-dimensional (2D) systems, altermagnetism can be induced by placing magnetic atoms away from the highest-symmetry Wyckoff position [20], arranging spin clusters with appropriate rotational relations [21], or using bilayer twisting [22,23], chemical modification [24,25], and stacking control [26-30]. These strategies have greatly expanded the material space of altermagnets, whose magnetic properties are predominantly understood in terms of spin vector configurations.

A more microscopic question is that once altermagnetic splitting is allowed, what selects whether the splitting has a *d*-wave, *g*-wave, or higher-harmonic angular form? Conventional symmetry descriptions usually treat magnetic sites as spin-bearing points. In realistic electronic systems, however, spin is carried by spatially extended wavefunctions whose angular structures are encoded in coordinate-dependent basis functions. These wavefunction geometries determine the directional dependence of electronic overlap and therefore provide a microscopic route of momentum-space spin splitting.

Here, we present a wavefunction-level framework for altermagnetism in 2D antiferromagnetic square (**sql**) lattices. Starting from symmetry-adapted polynomial wavefunctions, we present a hierarchy in which wavefunction geometry is manifested through local-crystal-field-selected orbital sectors, transferred to anisotropic hopping, and ultimately encoded in the harmonic form of altermagnetic spin splitting. Identical orbital sectors preserve conventional antiferromagnetic spin degeneracy, while intertwined orbital sectors generate altermagnetic splitting. In particular, intertwined linear and quadratic wavefunctions give rise to *d*-wave and *g*-wave altermagnetism, respectively. First-principles calculations on **mcm**-type material platforms and reticular candidates confirm the predicted high-symmetry-line degeneracy and generic-*k* splitting, consistent with intertwined quadratic wavefunctions and the resulting *g*-wave altermagnetic anisotropy. Our work establishes a link between symmetry-adapted wavefunctions, orbital manifestations, hopping anisotropy, and altermagnetic electronic structure.

*Wavefunction framework*. To describe this generally, we first formulate a wavefunction-level framework based on symmetry-adapted polynomial functions

$$\psi(\boldsymbol{r}) = \sum_n c_n f_n(\boldsymbol{r}), \tag{1}$$

where $\boldsymbol{r}$ denotes the spatial coordinate, $f_n(\boldsymbol{r})$ represents symmetry-adapted basis functions transforming according to irreducible representations of the underlying point-group symmetry, and $c_n$ is corresponding expansion coefficient. These basis functions encode the spatial geometry of the electronic wavefunctions and therefore determine the directional structure of electronic overlap.

* Y.C. and P.X. contributed equally to this work.

† Contact author: hflv@ustc.edu.cn

‡ Contact author: xjwu@ustc.edu.cn

Since electronic hopping originates from wavefunction interference, the hopping amplitude ($t$) between neighboring states separated by lattice vector $\boldsymbol{R}$ can generally be written as

$$t(\boldsymbol{R}) \sim \int f_i(\boldsymbol{r})\hat{H} f_i(\boldsymbol{r}-\boldsymbol{R})\mathrm{d}\boldsymbol{r}, \quad (2)$$

where $\hat{H}$ is the single-particle electronic Hamiltonian and $\boldsymbol{R}$ denotes the relative displacement between neighboring lattice sites. The hopping amplitude therefore directly inherits the geometry and angular symmetry of the underlying wavefunctions.

The corresponding momentum-space Hamiltonian ($H$) is obtained through Fourier transformation

$$H(\boldsymbol{k}) = \sum_{\boldsymbol{R}} t(\boldsymbol{R}) e^{\mathrm{i}\boldsymbol{k}\cdot\boldsymbol{R}} \quad (3)$$

where $\boldsymbol{k}$ is the crystal momentum in reciprocal space. Consequently, the angular anisotropy of the electronic structure becomes directly determined by the symmetry of the underlying wavefunctions. Within this framework, momentum-dependent spin splitting emerges as a direct consequence of symmetry-adapted wavefunction interference.

In realistic materials, the symmetry-adapted wavefunctions can be selected and stabilized by local crystal-field environments. The crystal field lowers the rotational symmetry of the electronic states and splits the wavefunction manifold into distinct angular components transforming under different irreducible representations. Consequently, specific local coordination geometries naturally favor particular polynomial wavefunctions and therefore determine the directional structure of electronic hopping.

For high-symmetry environments, the isotropic wavefunctions preserve equivalent hopping amplitudes along different spatial directions. In contrast, reduced crystal-field symmetry can generate anisotropic wavefunctions whose overlap depends strongly on direction, generating unequal hopping amplitudes and momentum-space anisotropy. The emergence of altermagnetic spin splitting can therefore be understood as a consequence of crystal-field-selected wavefunction geometry.

Within this framework, conventional atomic orbitals appear as physical realizations of the underlying polynomial wavefunctions. In 2D lattices, the symmetry-adapted wavefunctions reduce to in-plane polynomial functions, such as

$$1, x, y, xy, x^2 - y^2, \cdots \quad (4)$$

which correspond to isotropic, linear, and quadratic wavefunction geometries with distinct angular structures. In realistic electronic systems, these polynomial wavefunctions are physically realized through crystal-field-selected orbital states. The square (**sql**) lattice therefore provides a minimal platform in which the relation between wavefunction geometry and momentum-space spin splitting can be explicitly established.

*Orbital realization in square lattices*—As shown in FIG. 1(a), the **sql** lattice has $D_{4h}$ point group symmetry in a 4.4.4.4 tessellation. Placing $z$-orientated opposite spins at nearest-neighbor vertices yields opposite spin sublattices connected by a pure translation ($\tau$), resulting in a $p_p4/mm'm'$ magnetic layer group symmetry. This two-sublattice-based (spin-up (↑) and spin-down (↓)) **sql** lattice provides a pristine conventional antiferromagnetic configuration (see Supplemental Material, Sec. I [52]).

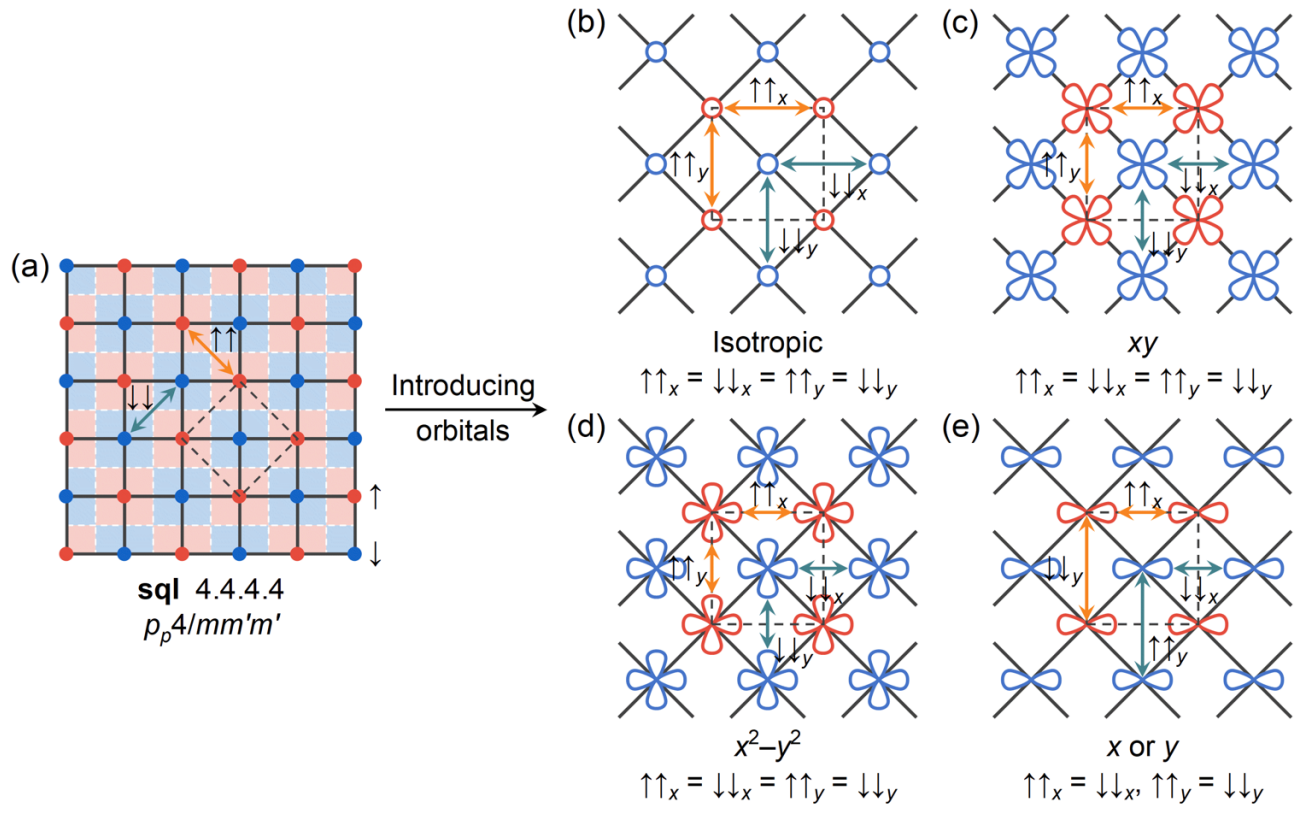

FIG. 1. Minimal model in a 2D square lattice with single-type orbitals. (a) Antiferromagnetic **sql** lattice with $p_p4/mm'm'$ magnetic layer group. (b-e) Top-view schematics of single-orbital scenarios for various orbitals. (b) Case of isotropic orbitals. (c) Case of quadratic $xy$ orbital. (d) Case of quadratic $x^2 - y^2$ orbital. (e) Case of linear ($x$ or $y$) orbitals. Red and blue symbols denote spin-up (↑) and spin-down (↓), respectively, while orange and green arrows denote hopping between ↑ states (↑↑) and between ↓ states (↓↓). Black dashed lines indicate unit cell considering opposite spins, and white dashed lines in (a) separate two spin sublattices by pure translation.

We next introduce orbital realizations of the symmetry-adapted wavefunctions. In the single-orbital limit, opposite-spin sublattices are described by identical orbital sectors, corresponding to the same underlying polynomial wavefunction geometry. FIG. 1(b)-1(e) illustrate the representative single-orbital configurations in the **sql** lattice.

For isotropic orbital states, such as 1, $z$, and $3z^2 - r^2$ [FIG. 1(b)], the electronic overlap remains equivalent along all in-plane directions. The corresponding wavefunction geometry therefore preserves full rotational equivalence between the two spin sectors, maintaining spin degeneracy throughout the Brillouin zone. A similar situation persists for quadratic orbitals such as $xy$ [FIG. 1(c)] and $x^2 - y^2$ [FIG. 1(d)], where these orbitals have wavefunction geometries ($D_{4h}$) compatible with the vertex symmetry. Consequently, the directional overlap remains symmetry equivalent between the two spin sectors, preventing spin splitting. For directional linear orbitals such as $x$ or $y$ [FIG. 1(e)], the identical spin-up (↑) and spin-down (↓) sectors preserve the $\{1'|\frac{1}{2}\frac{1}{2}\}$ generator in its magnetic layer group $c_pm'm'm$, preserving spin degeneracy across the entire Brillouin zone.

Consequently, to induce altermagnetism, the ↑ and ↓ sectors must be associated with distinct orbital configurations. FIG. 2(a) illustrates a minimal dual-orbital model with intertwined linear orbitals, where $\psi_\uparrow = x$ and $\psi_\downarrow = y$. In this configuration, the perpendicular orientation of $x$ and $y$ axes enforces $p4'/mm'm$ symmetry, with the fourfold rotation ($C_4$) axis shifted to $(\frac{1}{2}, 0)$. This **Lieb**-like geometry enables the emergence of $d$-wave altermagnetism (see Supplemental Material Sec. II [52]). This minimal model can be used to describe systems with a **Lieb** lattice, such as $V_2Se_2O$ and its family, where non-magnetic atoms anchor the spin density into $p_x$ and $p_y$ orbital shapes [19,20].

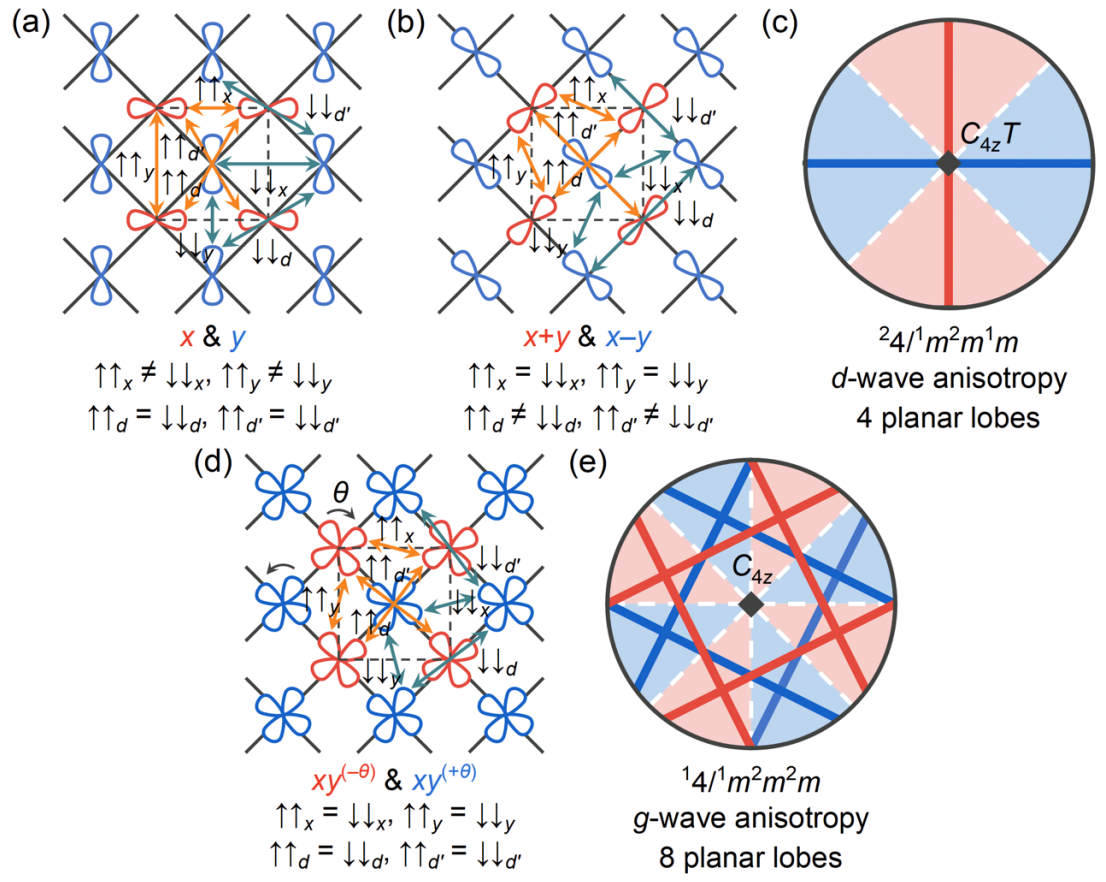

FIG. 2. Minimal model in a 2D square lattice with dual-type orbitals. (a) Minimal dual-orbital configuration with intertwined linear orbitals, producing $d$-wave altermagnetism. (b) A $\frac{\pi}{4}$-rotated version of (a). (c) Schematic illustration of the $d$-wave anisotropy in (a) and (b). (d) Minimal dual-orbital configuration with intertwined quadratic orbitals, featuring a linear combination of $xy$ and $x^2$–$y^2$ basis. (e) Schematic illustration of the $g$-wave anisotropy in (d).

FIG. 2(b) shows a related dual-orbital model with $\psi_\uparrow = x + y$ and $\psi_\downarrow = x - y$, which can be regarded as a $\frac{\pi}{4}$ rotation of FIG. 2(a). This configuration has a $p4'/mbm'$ symmetry, also characterized by $d$-wave altermagnetic anisotropy. Both intertwined linear-wavefunction configurations produce second-order momentum anisotropy corresponding to $d$-wave altermagnetism, with opposite spin sublattices connected by $C_{4z}T$ at the point group level [Fig. 2(c)].

We next consider intertwined quadratic orbitals derived from the $xy$ sector. As shown in FIG. 2(d), the ↑ and ↓ orbitals rotate clockwise ($-\theta$) and anticlockwise ($+\theta$) by a generic angle around the local $C_4$ axis, producing a linear combination of $xy$ and $x^2 - y^2$ components (see Supplemental Material, Sec. II [52]). The two orbital states are intertwined at the point group level, forming a $g$-wave anisotropy [FIG. 2(e)]. This configuration obtains the magnetic layer group of $p4/mbm$, which no longer contains $T$ symmetry. The resulting electronics structure exhibits spin degeneracy along all high-symmetry paths, including Γ(0,0), M($\pi$,$\pi$), X($\pi$,0), Δ($u$,0), Σ($u$,$u$), and Z($\pi$,$v$), where the corepresentations of these $k$-vectors are all doubly degenerate. Consequently, a fourth-order $k$-dependent spin-splitting emerges at general $k$-points D($u$,$v$) in the Brillouin zone with the form of $k_x k_y (k_x^2 - k_y^2)$ [31].

The wavefunction symmetry discussed above is directly reflected in the corresponding TB Hamiltonians (see Supplemental Materials, Sec. II [52]). For single-orbital configurations, the opposite-spin sectors remain related by $PT$ or $\tau T$ operations, and the resulting momentum-space Hamiltonian preserves an effective centrosymmetric structure. Consequently, the characteristic polynomial of the Hamiltonian factorizes into eigenvalues with double algebraic multiplicity, enforcing spin degeneracy throughout the Brillouin zone.

In contrast, intertwined dual-orbital configurations modify the symmetry relation between the two spin sectors and remove the effective centrosymmetric equivalence of the Hamiltonian. The characteristic polynomial no longer preserves globally repeated eigenvalues, allowing momentum-dependent lifting of spin degeneracy. For intertwined linear-wavefunction sectors, the resulting band splitting exhibits the characteristic $d$-wave anisotropy. For intertwined quadratic-wavefunction sectors, the symmetry constraints still enforce degeneracy along conventional high-symmetry paths, whereas the degeneracy is lifted at generic momentum points, producing higher-order $g$-wave anisotropy. The harmonic order of the momentum-space spin splitting is therefore directly inherited from the symmetry hierarchy of the underlying wavefunctions.

*Microscopic hopping mechanism*. We next elucidate the physical mechanism of spin-splitting from the perspective of orbital interactions and hopping. Given identical shapes of the square lattice in real space and reciprocal space, the electronic properties in momentum space can be directly inferred from the real-space orbital distribution. The hopping between same-spin states [↑↑ and ↓↓, FIG. 1(a)], arising from the spatial overlap of orbitals, determines whether spin degeneracy is preserved or lifted.

For a single-orbital system [FIG. 1(b)-(e)], the hopping amplitudes between same-spin states are equal in all directions, *e.g.*, $↑↑_x = ↓↓_x$ and $↑↑_y = ↓↓_y$ for $x$ and $y$ directions, respectively. This equality preserves Kramers degeneracy without SOC in the $D_{4h}$-symmetric system. Here, symbols such as $↑↑_x$ and $↓↓_y$ denote hopping channels between same-spin orbitals along the corresponding directions (*e.g.*, $x$ and $y$). Note that this notation is only used for qualitative description and does not correspond to explicit tight-binding parameters. Conversely, for a dual-orbital system (FIG. 2), orbital anisotropy leads to unequal hopping. For example, the combination of $\psi_\uparrow = x$ and $\psi_\downarrow = y$ in FIG. 2(a) results in asymmetric hopping along $x$ and $y$ directions ($↑↑_x \neq ↓↓_x$ and $↑↑_y \neq ↓↓_y$), while the hopping remains equal along the diagonal ($d$ and $d'$) directions ($↑↑_d = ↓↓_d$ and $↑↑_{d'} = ↓↓_{d'}$). This directly leads to spin-splitting along $k_x$ and $k_y$ paths but preserves spin degeneracy along $k_x \pm k_y$ paths. The relation between the anisotropic hopping amplitudes ($↑↑_x - ↓↓_x = ↓↓_y - ↑↑_y$) explicitly links the orbital hopping anisotropy to spin-splitting, ensuring equal but opposite splitting along $k_x$ and $k_y$ directions and forming the $d$-wave pattern.

A similar situation occurs for the rotated model in FIG. 2(b), where the diagonal hopping differs ($↑↑_d \neq ↓↓_d$ and $↑↑_{d'} \neq ↓↓_{d'}$) and the hopping along the $x$ and $y$ directions remains symmetric ($↑↑_x = ↓↓_x$ and $↑↑_y = ↓↓_y$). Consequently, spin-splitting occurs along $k_x \pm k_y$ directions but degeneracy is maintained along the $k_x$ and $k_y$ directions. For the intertwined $d$ orbital case in FIG. 2(d), the combination of quadratic basis results in $↑↑_x = ↓↓_x$, $↑↑_y = ↓↓_y$, $↑↑_d = ↓↓_d$, and $↑↑_{d'} = ↓↓_{d'}$, and the hopping is symmetric along all high-symmetry paths. However, at general $k$-points, the hopping components of ↑↑ and ↓↓ differ, resulting in the characteristic $k$-dependent spin polarizations of $g$-wave altermagnetism.

*Material candidates*. To demonstrate the proposed framework in realistic systems, we consider 2D materials with **mcm** topology, where crystal-field environments stabilize intertwined quadratic wavefunctions at magnetic centers (see methods in Supplemental Material, Sec. III [52]). A representative example is the recently identified bulk altermagnet $Nb_2FeB_2$ , which comprises the pentagonal $FeB_2$ layers separated by Nb interlayers [14,32]. The $FeB_2$ monolayer obtains $p4/mbm$ layer group symmetry, where magnetic Fe atoms occupy $2a$ position that forbids $d$-wave altermagnetism. Critically, the presence of nonmagnetic boron atoms anchors the Fe $3d$ orbitals into specific orientations, as evidenced by the spin density distribution, which aligns with our minimal model for $g$-wave anisotropy. Accordingly, the electronic band structure of the $FeB_2$ monolayer maintains spin degeneracy along all high-symmetry $k$-paths (FIG. S1, see Supplemental Material, Sec. IV [52]). Note that interchanging magnetic and nonmagnetic atoms to create an $M_2X$ configuration could enable $d$-wave altermagnetism by shifting the $C_{4z}$ axis away from the magnetic centers [20,33].

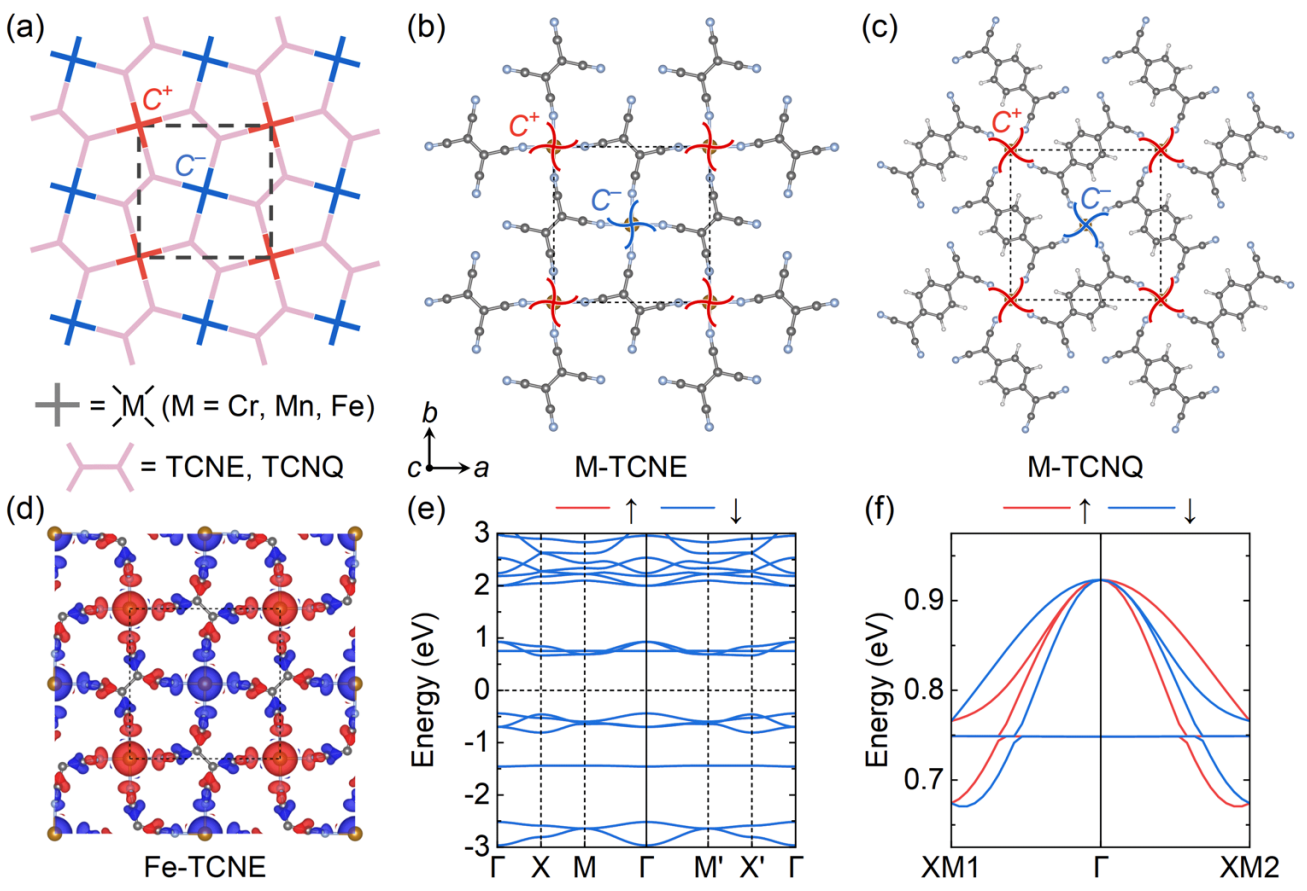

FIG. 3. MOF candidates M-TCNX (M = Cr, Mn, Fe; TCNX = TCNE, TCNQ). (a) Schematic of M-TCNX in an **mcm** lattice with each metal node (red and blue, fourfold symmetry) coordinated by alternately oriented ligands (pink, twofold symmetry). (b, c) Top view crystal structure of (b) M-TCNE and (c) M-TCNQ with opposite chirality ($\boldsymbol{C}^{+}$ for ↑ and $\boldsymbol{C}^{-}$ for ↓) at spin centers. (d) Spin density distribution of flat Fe-TCNE monolayer with the isosurface value of 0.001 e bohr$^{-3}$. (e, f) Spin-resolved band structure of flat Fe-TCNE along (e) high-symmetry directions and (f) general $k$-path $\mathrm{XM1}(\pi,\pi/2)\rightarrow\Gamma(0,0)\rightarrow\mathrm{XM2}(\pi,-\pi/2)$.

Building upon the **mcm** template, metal-organic frameworks (MOFs) provide more flexibility for realizing the proposed wavefunction hierarchy through tunable coordination environments and lattice topology [12,16,20,24,34-43]. MOFs are not solely introduced as isolated material examples, but as a platform that enables the implementation of the orbital-engineering strategy. Within this framework, the modular nature of MOFs enables controlled realization of the required orbital configurations and lattice symmetries. FIG. 3(a) illustrates the proposed 2D MOF structure of M-TCNX with an altermagnetic ground state within the **mcm** framework, where metals (M = Cr, Mn, and Fe) coordinate with tetracyanoethylene [TCNE, FIG. 3(b)] or 7,7,8,8-tetracyanoquinodimethane [TCNQ, FIG. 3(c)] ligands, demonstrating promising synthetic accessibility [44] (see Supplemental Material, Sec. V [52]). In flat M-TCNX with $D_{4h}$ crystal symmetry, each magnetic metal node maintains fourfold symmetry while being coordinated by alternately oriented ligands with twofold symmetry. This arrangement generates opposite chirality ($\boldsymbol{C}$) at opposing spin centers, *e.g.*, $\boldsymbol{C}^{+}$ for ↑ center and $\boldsymbol{C}^{-}$ for ↓ center, which manifests as opposite wavefunction chirality in the spin density distribution [FIG. 3(d)].

Among these candidates, flat Fe-TCNE emerges as a representative system based on its substantial spin-splitting magnitude and magnetic interactions [FIG. 3(e) and 3(f), TABLE S1]. The remaining flat monolayers show the same qualitative behavior, namely spin degeneracy along high-symmetry paths and splitting along a general $k$-path (FIG. S2). As shown in FIG. 4(f), flat Fe-TCNE reaches a maximum of 41 meV at the conduction band minimum (CBM) along the general $k$-path $\mathrm{XM1}(\pi,\pi/2)\rightarrow\Gamma(0,0)\rightarrow\mathrm{XM2}(\pi,-\pi/2)$ . This pronounced spin-splitting in M-TCNX mainly originates from hybridized $d_{z^2}$, $d_{xz}$, and $d_{yz}$ orbitals of the metal centers (FIG. S3). Under the chiral ligand field, their in-plane projections exhibit $xy^{(\theta)}$ characteristics, with $\theta$ = 44.18° and 8.67° for Fe-TCNE and Fe-TCNQ, respectively. Notably, SOC and easy-axis orientation have negligible affects on spin-splitting properties in M-TCNX systems, confirming that the splitting originates from symmetry-protected orbital-wavefunction anisotropy rather than relativistic effects (FIG. S4).

Under realistic structural relaxation conditions, initially flat M-TCNX layers undergo slight buckling distortions due to metal-center coordination and crystal field effects. To address this, we examine buckled M-TCNX monolayers with $D_{2d}$ symmetry that retain **mcm**-derived altermagnetism (TABLE S1 and FIG. S5). Phonon spectrum calculations confirm the dynamical structure stability of buckled Mn-TCNE and Fe-TCNE, which exhibit minimal imaginary frequencies. In contrast, free-standing buckled Cr-TCNE and M-TCNQ systems display imaginary optical branches under $D_{2d}$ symmetry, suggesting possible relaxation into slightly lower-symmetry configurations (FIG. S6). Importantly, these subtle distortions negligible impact the fundamental altermagnetic mechanism. Buckling preserves key electronic characteristics, including magnetic ground state, SOC-independent $g$-wave spin-splitting behavior (FIG. S7), and orbital anisotropy, demonstrating robustness of the altermagnetic state against structural corrugations. This resilience underscores the experimental feasibility of $g$-wave altermagnetism in MOF monolayers and validates the general applicability of our orbital engineering model framework.

*Conclusion*. In summary, we present a wavefunction-level framework for altermagnetism in 2D square lattices. Starting from symmetry-adapted polynomial functions, we show that the harmonic order of momentum-dependent spin splitting is directly connected to the geometry of the underlying wavefunctions. While identical orbital sectors preserve conventional antiferromagnetic degeneracy, intertwined linear and quadratic wavefunctions naturally generate $d$-wave and *g*-wave altermagnetism, respectively. The corresponding TB Hamiltonians and hopping anisotropy further reveal how momentum-space spin splitting emerges from inequivalent wavefunction sectors without spin-orbit coupling. Our framework establishes a direct hierarchy connecting symmetry-adapted wavefunctions, crystal-field-selected orbitals, momentum-space anisotropy, and realistic material platforms, offering a unified perspective for understanding and designing altermagnetic electronic structures through wavefunction perspective.

*Acknowledgements*. The research is supported by the National Natural Science Foundation of China (Grants Nos. 22225301, 22303092, and 22321001), the Quantum Science and Technology-National Science and Technology Major Project (Grant No. 2021ZD0303302), the Fundamental Research Funds for the Central Universities (Grants No. 20720250005, WK2490000001, and WK2490000002), the open research fund of Key Laboratory of Precision and Intelligent Chemistry, the robotic AI-Scientist platform of Chinese Academy of Sciences, and the Super Computer Center of USTCSCC and SCCAS.

*Note added*. Recent studies have highlighted the important role of orbital degrees of freedom in altermagnetism [45-50]. Furthermore, recent studies have demonstrated altermagnetic states in metal–organic framework (MOF) systems and explored their associated electronic properties [43,51]. Building on these developments, the present work provides a complementary wavefunction-level perspective that connects symmetry-adapted polynomial functions, crystal-field-selected orbital sectors, momentum-space anisotropy, and altermagnetic spin splitting within a unified hierarchy. Within this perspective, realistic material systems serve as physical realizations of the underlying wavefunction geometry and symmetry framework.

*Data availability*. The data are available from the authors upon reasonable request.